\pdfoutput=1  
\documentclass[11pt]{article}

\usepackage{preprint}

\usepackage[T1]{fontenc}
\usepackage[utf8]{inputenc}
\usepackage{lmodern}
\usepackage{textcomp}
\usepackage{microtype}
\usepackage{hanging}          

\usepackage{tcolorbox}
\tcbuselibrary{skins,breakable}
\newtcolorbox{recbox}{breakable,enhanced,colback=black!3,colframe=black!35,
  boxrule=0.4pt,left=10pt,right=10pt,top=8pt,bottom=8pt,arc=1pt}

\usepackage{xurl}
\PassOptionsToPackage{colorlinks=true,allcolors=black,citecolor=black,urlcolor=black}{hyperref}
\usepackage{hyperref}
\usepackage{orcidlink}
\title{Why Public Service AI Governance Frameworks Risk Failing in the
Age of General-Purpose AI:\\ Lessons from Policing}

\newcommand{\authorblock}{%
  \raggedright
  \textbf{Sam Relins}\textsuperscript{a,b,*}\,\orcidlink{0009-0001-7868-9835}\quad
  \textbf{Daniel Birks}\textsuperscript{a,b}\,\orcidlink{0000-0003-3055-7398}\par
  \vspace{0.8em}
  {\small
   \textsuperscript{a}\,ESRC Vulnerability and Policing Futures Research Centre\\
   \textsuperscript{b}\,School of Law, University of Leeds, Leeds, UK\\[0.5em]
   \textsuperscript{*}\,Corresponding author:
   \href{mailto:s.relins@leeds.ac.uk}{s.relins@leeds.ac.uk}\par}%
}
\author{\authorblock}

\headertitle{AI Governance in the Age of General-Purpose AI}

\begin{document}

\maketitle

\vspace{-0.7em}
{\footnotesize\itshape\raggedright
Preprint. This manuscript has been submitted for publication and has not
yet been peer reviewed.\par}
\vspace{0.6em}

\begin{abstract}
\noindent
Public services face growing pressure to adopt artificial intelligence
(AI) to close the gap between rising demand and falling resources. That
pressure has intensified with the arrival of general-purpose AI (GPAI):
AI built on large language models, that can be directed by prompt alone
to perform an effectively unbounded range of tasks. We argue that the
properties that make these models attractive: their generality,
accessibility, and the low cost and expertise required to deploy them,
undermine the conditions under which AI safety has historically been
pursued. The safety concepts that public service governance frameworks
foreground - accuracy, bias, explainability, and accountability - were
made tractable by narrow, purpose-built AI, and the mitigations that
guidance documents continue to prescribe presuppose exactly what GPAI
removes. Accuracy cannot be quantified over unbounded outputs. Bias
cannot be disaggregated when outputs are free-text judgements rather
than categorical predictions. Explainability gives way to the appearance
of explanation, and accountability erodes as outputs are optimized to
persuade. We develop this argument through the case of policing, where
the consequences of governance failure are among the most severe, and
show why the same failure is likely to recur across other public
services. The two mitigations that dominate policing AI strategy - expert evaluation and human-in-the-loop oversight - both rest on
assumptions that GPAI violates. The result is a structural inversion:
safety assurance shifts from an intrinsic feature of building an AI tool
to an optional, external add-on. We recommend a clear taxonomic
distinction between narrow and general-purpose AI in governance
documentation, a preference for technological parsimony, a pause on
operational deployment of GPAI in policing until adequate evidence
exists, and a coordinated national safety infrastructure with the
independence and authority to generate that evidence and determine when
responsible deployment is achievable.
\end{abstract}


\section{Introduction}

Public services across the United Kingdom face a familiar and
intensifying set of pressures: rising and diversifying demand, shrinking
budgets, and growing public expectations. Policing is no exception.
Police are expected to do more with less (Policing Productivity Review,
2023), and political leaders are increasingly drawn to artificial
intelligence as a means of squaring this circle. The appeal is
understandable. AI promises to automate routine administrative work,
surface insights from vast quantities of unstructured data, and support
decision-making in complex operational environments (Department for
Science, Innovation and Technology {[}DSIT{]}, 2025a). In a sector where
officers spend substantial portions of their time on paperwork and
information processing (Policing Productivity Review, 2023), the
prospect of reclaiming that time for frontline crime reduction and
community safety work is genuinely attractive.

AI is not a new proposition in policing. Some of the earliest AI
technologies to enter public consciousness - facial recognition,
recidivism prediction, predictive crime mapping - were policing
applications, though they often arrived as objects of concern rather
than enthusiasm (Angwin et al., 2016; Lum \& Isaac, 2016). The reasons
for that concern are well established. Conventional software can make
mistakes, but AI's mistakes can be harder to hold to account and faster
to propagate. The reasoning behind an output is often difficult to
interrogate (Burrell, 2016), and automation can carry a single flaw to
many people at once (Citron \& Pasquale, 2014). In policing, where
decisions bear on safety and liberty and rarely fall evenly across the
population, both properties carry particular weight.

These concerns have also been productive. Policing was among the first
domains in which society confronted high-stakes, rights-bearing
algorithmic decision-making. It was in that sense a proving ground: the
governance and assurance frameworks now relied upon across the public
sector were shaped in no small part by the effort to make algorithmic
policing accountable, explicable, and fair (Angwin et al., 2016;
Chouldechova, 2017; Harcourt, 2007; Oswald et al., 2018; \emph{R
(Bridges) v Chief Constable of South Wales Police,} 2020). That history
is precisely why those frameworks warrant scrutiny now. Recent
developments have accelerated the adoption of AI in policing while
raising serious questions about whether the safety evidence accumulated
to date applies to the technologies now being deployed (Bengio et al.,
2025).

General-purpose AI (GPAI) - built on large language models such as
GPT-4, Claude, Gemini, and Llama, and reaching the public through
systems such as ChatGPT - represents a paradigm shift in AI development
(Bengio et al., 2025; Bommasani et al., 2021). Earlier AI systems were
built to perform specific, circumscribed tasks, such as determining
whether two facial images match or estimating the likelihood that an
individual will reoffend. GPAI models are general by design. They accept
instructions in natural language and produce outputs that, in many
cases, replicate work that would previously have required specialist
human expertise. This has prompted a dramatic surge in adoption (Bengio
et al., 2025) and lent fresh momentum to the political enthusiasm,
described above, for AI across public services, policing included (DSIT,
2025a). It has also, we argue, marked a decisive departure from the
systems on which the existing safety literature was built.

A brief note on terminology. Following the EU AI Act (European Union,
2024, Arts. 3(63), 3(66); see also Bengio et al., 2025), the literature
distinguishes the GPAI \emph{model}: the trained artefact itself (GPT-4,
Claude, Gemini, Llama), from the GPAI \emph{system}: that model embedded
in an interface or application (ChatGPT, or a bespoke tool built upon
it). Our concern lies with the model: the properties that drive the
paradigm shift this paper addresses - unbounded task scope, open-ended
natural-language output, learned bias, fluent rationalization - are
dispositions of the model class. They travel with the model into every
system built upon it, from general-purpose products such as ChatGPT and
Claude to bespoke tools whose tailored packaging can give the appearance
of a narrow, purpose-built alternative. We accordingly take the model as
our unit of analysis, and use ``system,'' ``tool,'' and ``application'' in
their ordinary sense throughout.

This paper argues that the features which make general-purpose AI so
transformational - its generality, its accessibility, and the dramatic
reduction in the cost and expertise required to deploy it - fundamentally alter the conditions under which AI safety has
historically been pursued, and that current governance frameworks have
not kept pace with this shift. We trace the development of AI from
narrow, purpose-built tools and the evidence-based discipline of safety
research that grew up around them. We then argue that GPAI departs
sufficiently from those earlier systems that the established evidence
base can no longer support claims about the safety of technologies built
on it. We conclude with recommendations for how policing, and public
services more broadly, might navigate this transition without
sacrificing the principles of accountability, transparency, and public
trust on which their legitimacy depends.

\section{Narrow AI and the Foundations of AI Safety}

The applications most associated with AI in policing are largely
examples of narrow or purpose-built AI: systems that perform a single,
well-defined task. Unlike conventional software, which follows
explicitly programmed rules, these systems learn statistical patterns
from labelled training data - an approach broadly described as machine
learning. This enables the automation of tasks that have otherwise
resisted specification as deterministic rules: recognizing faces,
interpreting handwriting, predicting outcomes from complex combinations
of variables (Jordan \& Mitchell, 2015).

\subsection{Why AI raises distinctive safety concerns}

Machine learning models produce outputs as statistical estimates. A
conventional software system - a database query, a spreadsheet formula - follows predetermined rules, providing the same output every time it
receives the same input, and that output can be verified by tracing the
logic of the program code. A machine learning recidivism model, by
contrast, does not determine outright whether someone will reoffend; it
estimates a probability, derived from whatever regularities the
underlying algorithm has found in historical cases. This distinction,
between software that follows known rules and models that make uncertain
inferences from observed patterns, gives rise to the safety challenges
specific to AI. Across existing AI safety and ethics frameworks, those
challenges are consistently understood in terms of four core properties
(College of Policing, 2025; European Union, 2024; INTERPOL \& UNICRI,
2024; National Institute of Standards and Technology, 2023; National
Police Chiefs' Council, 2023; Organization for Economic Co-operation and
Development, 2019; Oswald et al., 2025):

\textbf{Accuracy:} The outputs of a machine learning model are
probabilistic estimates, and all such estimates carry some rate of
error. The safety question is not whether a model will sometimes be
wrong, but whether its error rate is known, in what direction errors
fall, and whether that rate is acceptable given the consequences of
being wrong in the context of deployment.

\textbf{Bias:} A model learns from the data it is given. If that data
reflects historical patterns of inequality, as criminal justice data
frequently does, the model may reproduce and potentially amplify those
patterns, not through any explicit instruction but as a statistical
artefact of its training.

\textbf{Explainability:} The internal workings of a trained model, which
are represented as potentially millions of numerical weights adjusted
during the learning process, do not correspond to human-readable rules
or reasoning. If a model flags someone as high-risk, it should be
possible, at some meaningful level, to understand what drove that
assessment.

\textbf{Accountability:} Where a decision was previously made by an
identifiable individual, the introduction of an AI system raises the
question of who bears responsibility for the decisions it informs or
displaces. A clear accountability structure must establish who is
answerable when the system's outputs lead to consequential action.

\subsection{How the narrow AI pipeline makes safety tractable}

Each of these properties has received sustained methodological attention
within machine learning research, and the resulting techniques form the
foundation of several core branches of contemporary AI safety work.

\textbf{Accuracy}: Developing a narrow AI system is, at its core, an
exercise in specifying and measuring correctness. The task is defined in
advance, labelled examples encode what a correct output looks like, and
a loss function makes the gap between predicted and desired output the
very quantity training seeks to minimize. Because the model operates
within a bounded input-output space, error rates are calculable and
failure cases identifiable (Japkowicz \& Shah, 2011). Accuracy in a
narrow AI system is not inferred post-hoc; it is what the model is
trained from the outset to optimize.

\textbf{Bias:} Because performance can be quantified, it can be
disaggregated: error rates can be broken down by ethnicity, gender, age,
or any other characteristic, and disparities measured directly (Barocas
\& Selbst, 2016; Chouldechova, 2017). The inputs are knowable and
modifiable - training data can be inspected for underrepresentation or
historical bias, and the learning process altered through reweighting,
resampling, or fairness constraints (Mehrabi et al., 2021). Bias in a
narrow AI system sits in identifiable components of a pipeline,
available to be audited and, in some cases, corrected before deployment.

\textbf{Explainability}: Although the internal workings of most machine
learning models do not correspond to human-readable rules, narrow AI
development offers concrete ways to address this. Because models operate
over defined inputs and produce a measurable output, it is possible to
ask which input features most influenced a given prediction - and
techniques exist to answer that question by systematically varying those
inputs (Lundberg \& Lee, 2017; Ribeiro et al., 2016). The problem can be
sidestepped entirely by choosing architectures interpretable by design
(Rudin, 2019). Importantly, inspecting feature influence also serves
development: it catches spurious correlations before deployment and
helps ensure that the model is learning genuine signal rather than
artefacts of the training data.

\textbf{Accountability}: Built for a single task, a narrow AI system
produces a bounded output - a risk score, a match probability, a
classification - and this clarity of function makes it straightforward
to specify where automated judgement ends and human responsibility
begins. Developers can document validated uses and known limitations, an
approach formalized in model cards (Mitchell et al., 2019), and
governance structures can assign clear lines of responsibility. The
narrowness of the tool, in other words, supports the narrowness of the
accountability question: it is possible to say, concretely, what the
system contributed to a decision and who is answerable for acting on
that contribution.

In each of these cases, the act of building a narrow AI tool: specifying
a task precisely enough to measure, inspecting training data, evaluating
against ground truth, iterating on errors - is itself the work that
makes safety questions tractable, whether or not a developer frames it
in those terms.

Where safety concerns do emerge, an evidence base has developed around
each application area. Recidivism prediction, predictive policing, and
facial recognition each emerge from circumscribed sub-disciplines with
their own empirical findings on bias, accuracy limits, and operational
performance. When systems perform the same core task, findings from one
deployment generalise to others, and auditing standards develop across
studies (Raji et al., 2020). Developers and procuring organisations
therefore inherit tools whose weaknesses are to some degree known, with
established evaluation protocols and a research community capable of
independently raising concerns.

Underlying all of this, narrow AI is genuinely difficult to build. It
requires labelled data of adequate quality, established methodology, and
specialist engineering expertise - each a bottleneck. The effect is not
that any system is guaranteed safe; the record of narrow AI in criminal
justice proves otherwise (Buolamwini \& Gebru, 2018; Dressel \& Farid,
2018; Richardson et al., 2019). It is that the pace and scope of
deployment stay bounded. When the pipeline from conception to deployment
is long, expensive, and technically demanding, operational systems
remain few enough, and each specific enough in function, that regulatory
scrutiny, institutional governance, and independent audit can
realistically keep pace.

\section{General-Purpose AI: A Different Paradigm}

General-purpose AI differs fundamentally from the narrow, purpose-built
systems that have until recently dominated discussions of AI in
policing. Where a narrow system is built to perform a specific,
pre-defined task, a GPAI model is not trained to perform any task in
particular: these models are trained on datasets of billions of
documents to model the statistical structure of language itself. The
capabilities that follow - summarization, translation, classification,
code generation, apparent reasoning, risk assessment - are not
engineered in but emerge from scale (Bommasani et al., 2021; Wei et al.,
2022). The result is a technology with no defined operational scope, in
which any task that can be described in natural language becomes
something the model will attempt.

The costs of building a narrow AI tool once defined the boundary of
practical deployment: only organizations with sufficient resources and
clearly scoped problems could justify the investment. GPAI changes that
calculus entirely. The same model that drafts an email can be prompted
to summarize a witness statement, triage a safeguarding referral, or
craft a complex intelligence analysis pipeline, and the marginal cost of
each new application is the cost of writing a prompt. The range of tasks
to which AI can plausibly be applied is no longer set by development
requirements but by the expressive range of natural language. The
benchmarks that frontier developers use to evaluate these models reflect
this directly. Where earlier systems were tested against narrow
task-specific metrics, current evaluations span legal, medical, and
scientific reasoning (Guha et al., 2023; Rein et al., 2023; Singhal et
al., 2025), mathematics (Glazer et al., 2024), software engineering
(Jimenez et al., 2024), and increasingly complex tasks involving
multi-step tool use (Mialon et al., 2024), with reported performance in
several domains approaching that of trained professionals (e.g., Rein et
al., 2023; Singhal et al., 2025).

Governments and public services have been quick to recognize this.
Nearly seventy countries have adopted national AI strategies
(Organization for Economic Co-operation and Development, 2024), an
increasing proportion of which are framed around the potential of
general-purpose AI. The UK's AI Opportunities Action Plan is
representative (DSIT, 2025a): an accompanying review puts the unrealized
savings from fully digitizing public services - with AI-driven
automation a central lever - at over \textsterling{}45 billion per year (DSIT, 2025b).
Policing frequently sits within this agenda. GPAI is being promoted as a
tool for drafting reports, triaging demand, summarizing intelligence,
and supporting frontline decision-making, with uptake driven as much by
bottom-up adoption among practitioners (Bright et al., 2024) as by
formal institutional rollout. The cumulative effect is that
general-purpose AI is increasingly positioned not as one tool among many
but as central to how the next generation of public services, policing
included, is expected to be reformed (DSIT, 2025a).

\section{The Collapse of Established Safety Assurances}

For most who have used tools such as ChatGPT, it is clear that GPAI does
indeed represent a genuine leap in capability. But the properties that
make GPAI general-purpose also make it much harder to assure as safe. On
the measures that matter most for responsible deployment - whether a
system can be trusted to be reliable, fair, interpretable, and subject
to meaningful oversight - GPAI introduces challenges qualitatively
different from those posed by narrow AI. To the best of our knowledge,
no adequate mitigations for these challenges have yet emerged.

The same four properties on which the previous section rested -- accuracy, bias, explainability, and accountability -- each invert as we move from narrow to general-purpose AI. Where narrow AI made accuracy calculable against a task fixed in advance, general-purpose AI offers no settled standard of correctness for its open-ended outputs. Where bias could be measured as a disparity in error rates and corrected, it becomes a structural property absorbed from the training corpus, including implicit associations that resist statistical detection. Where feature-attribution methods offered at least partial transparency, fluent rationalizations now supply only the appearance of explanation. And where a bounded output left a clear line at which human responsibility began, general-purpose AI automates the analytical work itself and erodes the oversight meant to contain it. We take each in turn.

\subsection{Accuracy as an Open Problem}

The challenge with narrow AI was one of degree. A system could be wrong,
and wrong in ways that were hard to anticipate, but what it meant to be
right was never in question - it had been settled in advance, when the
task was defined. An image classifier either names the correct category
or it does not; a risk score is either well calibrated or it is not.
General-purpose AI poses a different kind of problem. Its outputs are
often open-ended natural language - argument, analysis, judgement - and
there is no equivalent definition of correctness to check them against.
The same fact can be expressed in countless ways, all equally valid;
correct facts can be assembled through faulty reasoning; relevant
evidence can be quietly omitted or over-emphasized; and no single
element need be wrong for the whole to mislead. Judging whether an
output has succeeded presupposes some agreed standard of what success
is, and, for most of what we now ask these systems to do, that standard
does not exist.

In measurement terms, this is a problem of construct validity (Jacobs \&
Wallach, 2021). Measurement presupposes a construct specified precisely
enough to be operationalized, and no such specification is available
here. Whether an output is fit for use depends on its factual soundness,
completeness, framing, and relevance to the decision at hand, and the
weight given to each varies with the task and the user. These features
vary with context, resist reduction to a single accuracy score, and are
in part matters of judgement rather than fact (Wallach et al., 2025).
Nor is this a critic's objection; it is the position of the field
itself. The frontier laboratories describe evaluation of their own
models as unsolved (Anthropic, 2023; Ganguli et al., 2022), benchmarks
are acknowledged to capture only narrow slices of the behavior that
matters in use (Liang et al., 2023; Raji et al., 2021), and the UK AI
Safety Institute (since renamed the AI Security Institute) has stated
that evaluations cannot currently provide confident assurance that a
system is safe (AI Safety Institute, 2024). The depth of the problem is
best seen where measurement should be easiest. Rein (2025) tested GPAI
models on tasks chosen because success could be checked mechanically - writing program code with unit tests - and found that outputs passing
these checks were frequently still unfit for use. What can be measured,
in other words, does not vouch for what cannot.

Safety compounds the difficulty in three ways. The space of possible
harms is open-ended, so no aggregate test can be assumed to surface
them. Harms are context-dependent: behavior benign in one deployment may
be harmful in another. And harms are contested, since what counts as a
biased assessment or a misleading framing is itself a matter of dispute
(Ganguli et al., 2022). Put simply, a field that cannot establish
whether outputs are fit for use cannot certify that they are safe.

The implications for evidence accumulation follow directly. With narrow
AI, research accrued around each technology in ways that transferred to
deployment: the input space was bounded, the success criterion fixed,
and a finding about a system performing a given task in one setting said
something meaningful about the same system performing the same task in
another, because the task itself was stable. General-purpose AI offers
no such stability. In the absence of a settled standard of success,
findings from one operational context generalize poorly to the next. The
general safety literature establishes that failure modes such as
hallucination and bias exist as broad categories but says little about
how they will manifest in any particular application (Weidinger et al.,
2023). Even the most ambitious benchmarking efforts acknowledge covering
only a fraction of the relevant evaluation space (Liang et al., 2023).
The practical consequence for a deploying organization is stark: there
is no pre-deployment evidence base to consult, and no established
methodology by which one could be built. Each deployment must therefore
generate its own evidence, under operational conditions.

\subsection{Bias as a Structural Property}

With narrow AI, bias is at least legible: where model outputs diverge
systematically across demographic groups, the disparity can in principle
be detected, quantified, and challenged. General-purpose AI does not
work in this way. Trained on corpora representing a substantial
proportion of all digitized text, models absorb the full spectrum of
human prejudice and structural discrimination that text contains (Bender
et al., 2021; Gallegos et al., 2024). The result is not just a model
that disadvantages certain groups through a quantifiable disparity in
error rates, but one whose language itself may carry discriminatory
patterns - describing one ethnic group's behavior as ``aggressive'' where
another's would be characterized as ``distressed'', or consistently
framing certain communities in terms of risk rather than vulnerability
(Hofmann et al., 2024). These patterns surface in the texture of
continuous prose and detecting them requires careful qualitative
analysis rather than statistical comparison - analysis that may simply
not be practical at the scale of thousands of outputs that GPAI, by its
very nature, enables. Any such analysis also describes a single version
of a model. Changes to training data, fine-tuning, or alignment
procedures may alter these patterns in ways that cannot be anticipated
from the previous version, and there is no established basis for judging
how often the assessment would need to be repeated.

Recent work suggests that bias in GPAI may present itself not only in
specific outputs but features of the model's learned associations
themselves. Bai et al. (2025) find that LLMs which pass standard
explicit bias benchmarks nevertheless exhibit pervasive implicit biases - including associations between race and criminality, and race and
weapons - detectable only through indirect, psychologically inspired
measures, and producing measurable effects on downstream decisions. The
authors draw an analogy to humans who sincerely endorse egalitarian
values yet show bias on the Implicit Association Test: alignment
training appears to shape what the model says in response to direct
questioning while leaving the underlying associations largely intact. If
this is right, the standard tools for verifying that bias interventions
have succeeded - explicit benchmarks - are precisely the form of direct
questioning these findings call into question.

\subsection{The Explainability Paradox}

General-purpose AI is fundamentally beyond the reach of the
explainability techniques that offered at least partial transparency for
narrow AI systems. The scale of the underlying models - hundreds of
billions of parameters, trained on trillions of words - and the
unbounded nature of both their inputs and outputs render them
qualitatively more opaque even than the deep neural networks that
already presented significant interpretability challenges within the
narrow AI paradigm. Mechanistic interpretability research has made some
progress in identifying internal representations within large language
models (Bricken et al., 2023; Elhage et al., 2022), but these methods
remain highly computationally intensive, difficult to translate into
actionable insight, and far from yielding deployment-ready explanations - limitations the research itself acknowledges (Bereska \& Gavves,
2024). Explainability of GPAI is therefore better characterized as an
experimental research programme than an assurance mechanism - and for
proprietary commercial models, it is not available at all. An
organization deploying a system built on a proprietary model accessed
via an API has no access to the model's internal representations such
analysis would require, and there is little reason to expect this to
change: systematic evaluation of what major AI developers actually
disclose finds that transparency on model internals - including weights
and architectural details - is extremely low across the board (Bommasani
et al., 2023), and commercial vendors have strong incentives to keep it
that way.

The more consequential problem, however, is not merely that
general-purpose AI is harder to explain - it is that it produces outputs
which create a compelling illusion of explainability. Models can
generate fluent, structured, apparently reasoned justifications for
every output they produce, and to most users, these justifications are
indistinguishable from genuine reasoning. A growing body of research
demonstrates that this appearance is unreliable. Models can articulate
step-by-step reasoning that reads as coherent and principled, but Turpin
et al. (2023) show that such articulations frequently do not reflect the
factors that actually determined the output - the model may arrive at a
conclusion through one set of internal processes and generate a
rationalization that bears little causal relationship to the actual
computation. Arcuschin et al. (2025) find that this occurs not only
under experimental conditions but in realistic deployment, with
production models generating such rationalizations at measurable rates.
Barez et al. (2025) conclude on this basis that articulated model
reasoning should not be treated as evidence of interpretability at all.

For governance frameworks that rely on explainability as the safeguard
through which human oversight is exercised and accountability
maintained, this presents a fundamental challenge. The ``explanations''
that GPAI provides are not windows into its reasoning but artefacts of
the same generative process that produces its substantive outputs,
subject to the same risks of hallucination, bias, and confabulation. A
policing professional reviewing a model-generated risk summary has no
way of knowing whether the accompanying justification reflects the
model's actual basis for its assessment or is simply the most
plausible-sounding narrative the model could construct after the fact.

\subsection{Accountability and the Erosion of Oversight}

The observation that GPAI can produce persuasive but misleading
justifications for its decisions changes what accountability can
meaningfully look like. A narrow AI output - a risk score, a
classification, a flag - presents itself as something to be evaluated.
General-purpose AI outputs do not. Extended natural-language analyses,
recommendations, and summaries blend with the kind of communication that
normally flows between human colleagues, in prose that is, by the nature
of these models, constructed to be coherent, well-argued, and
compelling. How these outputs should be assessed, and against what
standard, remains a fundamental but wholly open question.

This is not a property that can be designed around. The fluency that
makes models useful is inseparable from what makes their outputs hard to
resist. Put another way, persuasive capability is a function of the
architecture, not an add-on. Hackenburg et al. (2025), in a large-scale
study deploying 19 language models across three experiments with nearly
77,000 participants, find that where post-training and prompting
interventions increase model persuasiveness, they do so at a systematic
cost to factual accuracy. The persuasive capacity these interventions
exploit is not one they introduce; it is latent in the model, activated
by choices that are difficult for deploying organizations to monitor or
constrain. An officer reviewing a model-generated assessment is
therefore not in a position analogous to reviewing a colleague's
analysis, where the quality of the reasoning is at least some guide to
the quality of the conclusion. They are reviewing an output whose
persuasive qualities are structurally independent of, and may be
inversely related to, its reliability.

\subsection{Proliferation and the Limits of Scrutiny}

The failures examined so far - unmeasurable accuracy, structural bias,
illusory explanation, eroded oversight - are each a property of the
technology itself. But the deeper problem is structural. As Section 2
established, building a narrow AI system meant engaging with the
questions on which safety depends, because task boundaries, data
provenance, and measurable behavior had to be constructed for the system
to be developed, and were therefore in principle inspectable.
General-purpose AI severs this connection. Its central proposition is
that such construction is no longer required - that a single
general-purpose model, wrapped in an interface, will perform the task
without the deploying organization needing to characterize it, measure
it, or perhaps even understand it. Engagement with safety-relevant
questions, once integral to building an AI tool, becomes discretionary,
layered on top of a deployment that is already complete. The danger is
not merely that answers are deferred but that the questions cease to be
legible: what the model has learned, how it decomposes the task, and
where its failures will cluster are obscured by the generality that
makes it compelling.

An organization that does pursue safety finds none of the resources
narrow AI could assume. The evidence base around predictive policing
accumulated because each task was circumscribed; no comparable
accumulation exists for GPAI, and the general-purpose safety literature
that has emerged in its place - alignment, emergent capability,
catastrophic risk - offers limited purchase on the operational questions
a specific deployment raises (Weidinger et al., 2023). To engage
seriously with safety here is not to apply established methods but to
attempt original research in a new and rapidly changing field. And the
space in which this problem arises has no boundary: the marginal cost of
a new application is the cost of writing a prompt, and the expertise
once required even to attempt deployment is no longer needed. The
promise of general-purpose AI is therefore inseparable from the collapse
of the conditions under which safety was previously achievable - not as
a side-effect that better design could restore, but as a direct
consequence of the generality, accessibility, and low cost that define
the paradigm.

\section{The Inadequacy of Prescribed Mitigations}

Current AI governance frameworks in policing acknowledge, to varying
degrees, that GPAI introduces new challenges. Some documents engage
substantively with these issues (College of Policing, 2025; European
Union, 2024; Oswald et al., 2025); others treat general-purpose AI as a
minor variation on a familiar theme (INTERPOL \& UNICRI, 2024; National
Institute of Standards and Technology, 2023). But even the more
thoughtful frameworks can currently only recommend the following
mitigations, which we argue are flawed.

\subsection{Expert Evaluation}

The most straightforward response to uncertainty about an AI system's
behavior is to require that qualified experts evaluate it before
deployment. For narrow AI, this is a coherent requirement. A substantial
community of ML practitioners has developed experience working through
the safety challenges that purpose-built systems present - characterizing failure, applying established evaluation methods, and
drawing on a body of prior findings that transfer across deployment
contexts. Such evaluation is demanding and often under-resourced, but it
is a legible activity, with an established body of knowledge organized
around specific model classes and applications and a population of
practitioners capable of applying it.

For general-purpose AI, neither exists. An expert can only assess a
system against criteria, using methods, informed by evidence. Section 4
established that for general-purpose AI there are no agreed criteria, no
established methodology, and no accumulated body of findings. Nor is
there a population of practitioners with meaningful experience of
resolving these questions in deployment, because the questions
themselves remain open at the frontier of the field. To call for expert
evaluation under these conditions is not to invoke a familiar safety
mechanism; it is to invoke expertise that does not yet exist, and to
treat that invocation as a solution.

\subsection{Human-in-the-Loop}

The most common safety mitigation proposed for AI in high-stakes
contexts is human review, validation, or approval of AI outputs before
they are acted upon. Sections 4.3 and 4.4 have already shown why GPAI
erodes human oversight in practice, which alone should trouble
proponents of this safeguard. At a more basic level, though, the
human-in-the-loop principle is conceptually incoherent when applied to
many of the tasks GPAI is being proposed to perform.

For narrow AI, oversight is workable because the tool is built for a
role the human is capable of overseeing. The AI is trained on a specific
dataset and has been evaluated to a known accuracy on a well-defined
task; its output is a discrete score or classification. The human does
something different: they bring in contextual information the model does
not have access to, they weigh the tool's reported strengths and
weaknesses, they apply professional judgement, and they assume
responsibility for the decision. This division of labor works because
the human is not being asked to check the AI's workings, only to decide
whether its output makes sense in the wider context they are aware of.

For general-purpose AI this division collapses, because the system is no
longer performing a circumscribed technical task alongside the human: it
is automating the human's analytical process itself. With document
synthesis, summarization, or narrative assessment, the output is not a
score but an extended, fluent, apparently well-reasoned analysis: the
kind the analyst would otherwise have produced. Identifying an error in
it requires the reviewer to already know, or to independently derive,
the correct analysis. But if they could do that, the system's
contribution would be redundant; if they cannot, meaningful verification
means replicating the very work the system was introduced to perform.

Human-in-the-loop oversight, in its current formulation, treats two
things as jointly achievable: the productivity benefits of automating
analytical work, and the safety benefits of human judgement over that
same work. For general-purpose AI, this is incoherent. The degree of
automation these tools promise is achievable only to the extent that
human oversight is relaxed. Meaningful oversight, conversely, demands
precisely the time, attention, and expertise that adoption was supposed
to free up.

\subsection{The Problem of ``High Stakes''}

Many frameworks prescribe enhanced scrutiny, or avoiding GPAI solutions
entirely, in ``high-stakes'' contexts, implying that lower-stakes uses
require less rigorous oversight. This distinction is more difficult to
maintain in practice than it appears in policy, particularly when
considering AI applications in public services generally and policing
specifically.

Who determines what constitutes a high-stakes use? In a large, complex
organization such as a police force, the same piece of AI-generated
output might inform decisions at multiple levels and in multiple
contexts. A summary of an incident report might seem like a low-stakes
administrative convenience, until it is the summary that a supervisor
reads when deciding whether to escalate a safeguarding concern. An
AI-generated triage of incoming reports might appear routine, until the
triage logic systematically deprioritizes a category of crime that
disproportionately affects a particular population. The stakes of a
given application are not fixed properties of the tool; they are
emergent properties of the organizational context in which the tool's
outputs circulate.

Furthermore, the cumulative effect of many ``low-stakes'' AI applications
may be far more consequential than any single high-stakes deployment. If
every officer in a force is routinely using AI in ``low-stakes'' contexts - to draft reports, summarize evidence, and process information - the
aggregate influence of these tools on the organization's knowledge base,
decision-making, and institutional memory may be profound, and
profoundly difficult to audit. The distinction between high-stakes and
low-stakes use cases, while intuitive, may offer false comfort by
directing scrutiny towards a small number of visible applications while
the broader fabric of organizational decision-making is quietly
transformed. It is not unreasonable to argue that policing is, in and of
itself, a high-stakes domain, and that governance frameworks should
proceed on that basis, rather than inviting protracted and ultimately
unresolvable debates about where exactly the threshold of concern should
fall.

\section{Recommendations}

The preceding analysis establishes that general-purpose AI differs from
narrow AI in ways that are structurally consequential for safety, and
that the governance frameworks currently guiding AI adoption in policing
have not reckoned with this difference. We argue that the
recommendations that follow are essential requirements for any
governance framework seeking to address the challenges GPAI poses in
policing. They are intentionally ambitious, reflecting the scale and
significance of those challenges, while recognizing the practical
constraints faced by individual policing organizations and national
bodies. The recommendations are cumulative: each depends upon those that
precede it.

\subsection{A Taxonomic Distinction in Governance}

\begin{recbox}
\noindent\textbf{Recommendation.} Governance frameworks, strategic documents, and
operational guidance addressing AI in policing should distinguish
explicitly between narrow, purpose-built AI and general-purpose AI, and
should specify which of their provisions apply to each.
\end{recbox}

Narrow AI and GPAI do not belong to the same technological category:
they differ in how they are built, in the failure modes they exhibit at
deployment, and in the assurance practices that can meaningfully be
applied to them. The distinction must follow the model, not the wrapper - a system built on a GPAI model is a GPAI system however narrow its
packaged task appears. The same applies to composition. Where a GPAI
model performs any step in a processing chain - extracting fields from
free text, drafting a summary that a later stage acts on, generating
features for a downstream classifier - the properties of that step
propagate to the output, and the system as a whole should be treated as
GPAI for assurance purposes. Documents that treat ``AI'' as a single
category implicitly suggest that approaches viable in one paradigm
transfer to the other; the argument of this paper is that they do not.

\subsection{Technological Parsimony as a Rule}

\begin{recbox}
\noindent\textbf{Recommendation.} Where a problem can be addressed by a narrow
tool, and the data, methodology, and expertise to build one are
available, that route should be preferred.
\end{recbox}

The corollary preference for narrow tools follows from the same logic.
The safety properties of narrow AI are in significant part a consequence
of the conditions under which it is built - bounded tasks, defined data,
measurable outputs, an accumulated domain literature. Those conditions
are themselves an asset, and forgoing them where they are available is a
choice rather than a default. Narrow tools come with a development
pipeline that makes assurance tractable; general-purpose AI does not.
Critically, decades of research and operational experience have produced
empirically validated best practices for the safe deployment of narrow
tools (Mitchell et al., 2019; Raji et al., 2020).

This principle of parsimony does not reject innovation, but resists
defaulting to GPAI simply because it requires no bespoke development.
Before adopting a general-purpose system, agencies should evaluate
whether a viable narrow-AI alternative could serve the operational need.
In many cases, the answer will be no - the required training data may be
unavailable, the task too novel, or development costs prohibitive. But
where narrow tools are viable, they should be preferred. Where they are
not, that determination should be explicit and documented, ensuring GPAI
is adopted because it is necessary, not merely convenient.

\subsection{A Pause on Operational Deployment}

\begin{recbox}
\noindent\textbf{Recommendation.} Operational deployment of general-purpose AI
models in policing should be paused until domain-specific evaluation
methodologies and empirical findings exist that are sufficient to
characterize their behavior in the contexts where they would be used.
\end{recbox}

The preceding sections have argued that the failure modes of GPAI are
consequences of the technology's architecture rather than incidental
properties amenable to procedural mitigation. The domain-specific
methodology needed to assess and mitigate those failure modes - to
understand how they manifest in particular policing tasks, to determine
whether outputs are fit for specific operational purposes, to identify
conditions under which deployment risks become unacceptable - does not
yet exist. To deploy in its absence is to proceed without the capacity
to know what one is deploying.

It may be argued that uncertainty itself is the baseline condition of
any AI application, and that governance frameworks have always managed
it through procedural safeguards. That premise is sound for narrow AI. A
risk score or a facial match is uncertain, but its uncertainty is
characterized: it comes with an error rate, a direction of known
failures, and a research literature that has documented how it degrades.
The uncertainty of a GPAI output in an operational policing context is
not uncertainty of that kind. It is an absence not merely of answers but
of the frameworks and methods that would make the relevant questions
legible.

There is also a timing argument. The conditions that would make
responsible deployment possible - evaluation methodologies, empirical
findings on operational behavior, professional expertise capable of
interpreting them - can only be built if they are treated as
prerequisites of deployment rather than workstreams to be pursued
alongside it. Once general-purpose AI is embedded in operational
practice, the institutional pressures that make a retreat from deployed
technology difficult - sunk costs, staff dependence, public commitments
attached to productivity projections - apply in full. In this context,
the potential pace of development with GPAI is deeply concerning. As
established above, the range of tasks to which GPAI can be applied is
bounded only by the expressive range of natural language, and the
marginal cost of attempting any one of them is the cost of writing a
prompt. Under these conditions, there is no realistic trajectory on
which evaluation infrastructure, built reactively in the wake of
deployment decisions, catches up.

The recommendation is therefore specified in terms of what would end the
pause. Domain-specific evaluation methodologies and empirical findings
sufficient to characterize GPAI behavior in specific policing contexts
are a concrete and achievable standard. Reaching it requires coordinated
investment in safety research, development of evaluation protocols
appropriate to specific operational tasks, and accumulation of findings
sufficient to support generalizable conclusions. That work is feasible;
it is also work that the pressure of prior deployment commitments will
tend to crowd out.

\subsection{Centralized Safety Infrastructure}

\begin{recbox}
\noindent\textbf{Recommendation.} An independent body with regulatory authority
over general-purpose AI use in policing should be established, with
responsibility for developing the evidence base required for responsible
deployment and determining when it is sufficient.
\end{recbox}

The conditions specified by the previous recommendation - domain-specific evaluation methodologies, empirical findings sufficient
to characterize GPAI behavior in specific operational contexts,
professional expertise capable of interpreting them - do not produce
themselves. They are the product of a sustained programme of work,
conducted under conditions that make the resulting knowledge cumulative,
public, and authoritative. Those conditions in turn require an
institutional structure adequate to deliver them. The features set out
below are not exhaustive, but each is necessary; in their absence, the
safety regime envisaged by 6.3 cannot be realized.

The body must have concentrated technical expertise. The capacity to
evaluate general-purpose AI in operational settings - to design
experiments and evaluations, distinguish robust findings from artefacts
of test construction, and trace the implications of model behavior for
specific tasks - is among the scarcest in the field, and cannot be
replicated across the many police forces, vendors, advisory boards, and
isolated academic groups. It must be pooled in a body of sufficient
standing to attract it and sufficient continuity to accumulate knowledge
over time. Distributing the function across the existing landscape,
where most actors cannot scrutinize vendor claims at all, is not an
alternative: it is the present situation, and the present situation is
what motivates this paper.

The body must also have independence from the interests that would
otherwise shape its findings. The incentive structure surrounding
adoption is particularly adverse to honest assessment: vendors have
commercial reasons to characterize their products as fit for purpose;
police organizations face productivity pressures and sunk costs that
make a finding of unfitness difficult to act on; governments attach
policy commitments to AI-driven efficiency gains. Without an actor with
the standing to say no, the path of least resistance is a performative
regime that ``safety-washes'' decisions taken on other grounds.
Independence must therefore extend not only to vendors but to policing
organizations and the governmental machinery that has staked political
capital on the productivity case. The criterion is demanding because the
alternative is a form of capture.

The body must have regulatory authority. Setting standards is not
enough: it must have the power to determine which applications may be
deployed, under what conditions, and when the evidence is sufficient to
lift the pause specified by 6.3. Without enforcement, those conditions
reduce to a recommendation that operational pressures can route around
force by force. That authority cannot end at the point of deployment.
Because a model's behavior can change with any revision to its training,
fine-tuning, or alignment, so an authorization issued against one
version of a model says nothing about its successors; the body must
therefore retain the power to impose conditions on deployed systems,
require re-evaluation when models change, and suspend authorizations
where the evidence supporting them no longer holds.

The body's work must also be transparent. Concentrating expertise in a
single institution answers the scarcity problem but introduces one of
its own: the body that evaluates general-purpose AI risks becoming the
only body positioned to evaluate those evaluations. The check is
transparency - methodologies, findings, and reasoning made available in
detail sufficient for substantive scrutiny by outside parties:
researchers, civil society, journalists, and external policing
professionals. This serves two ends. It enables independent contribution
to an evidence base too large for any one institution to internalize,
and it underpins democratic accountability: policing exercises coercive
authority on the public's behalf, and the technologies mediating that
authority cannot be governed by determinations the public cannot
examine. Without transparency, the concentration of expertise and
authority the preceding requirements demand becomes a different kind of
opacity.

This is a demanding recommendation, and we state it in that form because
the analysis leads here. It calls for institutional capacity that does
not currently exist in the configuration described, in any jurisdiction,
and it places that capacity in advance of the deployment activity it is
intended to govern. It is not, however, without precedent. The European
Union's AI Act (European Union, 2024) designates law enforcement uses of
AI as high-risk and conditions their deployment on conformity assessment
rather than presuming it permissible by default. Singapore's Model AI
Governance Framework (Personal Data Protection Commission \& Infocomm
Media Development Authority, 2020) and AI Verify initiative (Infocomm
Media Development Authority \& Personal Data Protection Commission,
2022) concentrate technical evaluation capacity in bodies with the
standing to set authoritative standards. National AI safety institutes,
now established in several jurisdictions and increasingly coordinated
internationally, demonstrate that concentrated public-sector expertise
in general-purpose AI evaluation is institutionally feasible. Police.AI,
recently established in the United Kingdom as a national capability for
AI in policing (Home Office, 2026), reflects a recognition that the task
requires coordinated national capacity rather than fragmented
force-level adoption. None combines the features described at the right
scale, but each demonstrates that institutional arrangements of this
general shape are within the realm of contemporary policy.

\section{Conclusion}

The pressure on policing, and public services more generally, to adopt
general-purpose AI is real, and the potential benefits are genuine. We
write from a point early enough in~policing's~adoption that the
important decisions~remain~open, but late enough that they are already
being taken. This is why the argument is worth making now, before those
decisions harden into investment and infrastructure. General-purpose AI
offers the prospect of dramatic productivity gains in a sector where
officers spend substantial proportions of their time on information
processing, report writing, and administrative tasks that consume
resources desperately needed elsewhere (Policing Productivity Review,
2023). Beyond automating routine tasks, the same technology could
meaningfully enhance the analytical functions that policing often lacks
the capability to perform at scale: identifying hidden patterns across
large volumes of administrative data or evidentiary capture, flagging
complex safeguarding concerns, or enabling more systematic use of
evidence in operational decision-making. An organization asked to do
more with less will naturally be drawn to a technology that appears to
offer exactly that. We believe general-purpose AI will be adopted across
public services, including policing, and the pressures driving that
adoption are real. We do not argue that this interest is misplaced, nor
make a case for rejecting the technology outright.

We do, however, argue that the conditions for responsible deployment of
GPAI do not currently exist, and that the pace of adoption risks
outstripping any effort to establish them. The question this paper
presses is one of sequence:~whether the evidence needed to
deploy~responsibly~is~established~before deployment or assembled after
it. The problem is structural: the qualities that make these systems
easy to deploy - generality, accessibility, low cost - are in large part
the absence of the development constraints that once forced adopters to
confront accuracy, bias, and accountability as a condition of use.
Current governance frameworks, built for a world of few, bounded tools,
prescribe mitigations that survive in form but have lost their
substance.

The risk, if the gap is not addressed, follows from the combination the
analysis has identified: systems whose outputs resist verification,
deployed where decisions affect liberty, safety, and welfare, overseen
by structures unable to detect failure - exposing policing to risk at
every scale. A hallucinated detail in a safeguarding summary that leaves
a child unprotected or draws a family into an intervention that was
never warranted. A systematically biased triage process that
deprioritizes an entire category of victim. A confidently reasoned
analytical output, wrong in substance but compelling in form, that
shapes a consequential operational decision. These acute risks sit
alongside the slower but potentially deeper problem of cumulative
effect: the gradual embedding of poorly understood systems into the
routine functioning of an institution, quietly reshaping the quality of
organizational knowledge, the independence of professional judgement,
and the capacity of oversight structures to detect and correct error - a
transformation that may become visible only once it is too deeply
entrenched to reverse easily.

These costs are not confined to the organization that incurs them.
Policing depends on public confidence to a degree that few other public
services do, and in some jurisdictions that confidence is already
strained. A well-publicized failure of the kind described above does not
stay with the case, the unit, or the agency that produced it. It is
inherited by everyone doing similar work - and the more visible the
failure, the wider the inheritance. The fact that the technology is
convenient and the political pressure to adopt is intense does not make
these problems less real.

The recommendations set out here - a taxonomic distinction between
narrow and general-purpose AI, a presumption in favor of narrow
purpose-built tools where they are viable, an independently produced
evidence base as a precondition for deployment in specific use cases,
and centralized safety infrastructure with the authority to build it and
enforce its conclusions - are demanding. They require investment,
institutional reform, and a willingness to resist the pressure to deploy
ahead of the knowledge needed to deploy well. But the alternative is to
let a transformative and poorly understood technology be driven by
commercial availability and operational convenience, with safety
addressed retrospectively, if at all. The opportunity GPAI represents is
real. Whether it is realized or squandered depends on whether policing
insists that the conditions for responsible use are met before, rather
than after, these systems become too deeply embedded to govern.

\section*{Acknowledgements}

The support of the Economic and Social Research Council (ESRC) is
gratefully acknowledged. Grant reference number: ES/W002248/1.

Claude (Anthropic) and ChatGPT (OpenAI) were used to copy-edit
author-written text and to provide adversarial review of the
manuscript's argument and expression. No substantive content was
generated by the tools; all edits and responses to review points were
reviewed, verified, and approved by the authors, who take full
responsibility for the content of the manuscript.

\section*{Conflict of Interest}

The authors declare that they have no conflict of interest.

\section*{References}
\addcontentsline{toc}{section}{References}
\begin{hangparas}{1.4em}{1}
\small
\setlength{\parskip}{0.6em}
AI Safety Institute. (2024). AI Safety Institute approach to
evaluations. Department for Science, Innovation and Technology.
\url{https://www.gov.uk/government/publications/ai-safety-institute-approach-to-evaluations}

Angwin, J., Larson, J., Mattu, S., \& Kirchner, L. (2016, May 23).
Machine bias: There's software used across the country to predict future
criminals. And it's biased against Blacks. \emph{ProPublica}.
\url{https://www.propublica.org/article/machine-bias-risk-assessments-in-criminal-sentencing}

Anthropic. (2023). Challenges in evaluating AI systems.
\url{https://www.anthropic.com/research/evaluating-ai-systems}

Arcuschin, I., Janiak, J., Krzyzanowski, R., Rajamanoharan, S., Nanda,
N., \& Conmy, A. (2025). \emph{Chain-of-thought reasoning in the wild is
not always faithful} {[}Preprint{]}. arXiv.
\url{https://arxiv.org/abs/2503.08679}

Bai, X., Wang, A., Sucholutsky, I., \& Griffiths, T. L. (2025).
Explicitly unbiased large language models still form biased
associations. \emph{Proceedings of the National Academy of Sciences,
122}(8), Article e2416228122.
\url{https://doi.org/10.1073/pnas.2416228122}

Barez, F., Wu, T.-Y., Arcuschin, I., Lan, M., Wang, V., Siegel, N.,
Collignon, N., Neo, C., Lee, I., Paren, A., Bibi, A., Trager, R.,
Fornasiere, D., Yan, J., Elazar, Y., \& Bengio, Y. (2025).
\emph{Chain-of-thought is not explainability} {[}Preprint{]}. AI
Governance Initiative, University of Oxford.
\url{https://aigi.ox.ac.uk/wp-content/uploads/2025/07/Cot_Is_Not_Explainability.pdf}

Barocas, S., \& Selbst, A. D. (2016). Big data's disparate impact.
\emph{California Law Review, 104}(3), 671--732.
\url{https://doi.org/10.15779/Z38BG31}

Bender, E. M., Gebru, T., McMillan-Major, A., \& Shmitchell, S. (2021).
On the dangers of stochastic parrots: Can language models be too big? In
\emph{Proceedings of the 2021 ACM Conference on Fairness,
Accountability, and Transparency} (pp. 610--623). Association for
Computing Machinery. \url{https://doi.org/10.1145/3442188.3445922}

Bengio, Y., Mindermann, S., Privitera, D., Besiroglu, T., Bommasani, R.,
Casper, S., Choi, Y., Fox, P., Garfinkel, B., Goldfarb, D., Heidari, H.,
Ho, A., Kapoor, S., Khalatbari, L., Longpre, S., Manning, S., Mavroudis,
V., Mazeika, M., Michael, J., \ldots{} Zeng, Y. (2025).
\emph{International AI Safety Report} (DSIT 2025/001). UK Department for
Science, Innovation and Technology.
\url{https://arxiv.org/abs/2501.17805}

Bereska, L., \& Gavves, E. (2024). Mechanistic interpretability for AI
safety---A review. \emph{Transactions on Machine Learning Research}.
\url{https://openreview.net/forum?id=ePUVetPKu6}

Bommasani, R., Hudson, D. A., Adeli, E., Altman, R., Arora, S., von Arx,
S., Bernstein, M. S., Bohg, J., Bosselut, A., Brunskill, E.,
Brynjolfsson, E., Buch, S., Card, D., Castellon, R., Chatterji, N.,
Chen, A., Creel, K., Davis, J. Q., Demszky, D., \ldots{} Liang, P.
(2021). \emph{On the opportunities and risks of foundation models}.
arXiv. \url{https://arxiv.org/abs/2108.07258}

Bommasani, R., Klyman, K., Longpre, S., Kapoor, S., Maslej, N., Xiong,
B., Zhang, D., \& Liang, P. (2023). \emph{The Foundation Model
Transparency Index} {[}Preprint{]}. arXiv.
\url{https://arxiv.org/abs/2310.12941}

Bricken, T., Templeton, A., Batson, J., Chen, B., Jermyn, A., Conerly,
T., Turner, N., Anil, C., Denison, C., Askell, A., Lasenby, R., Wu, Y.,
Kravec, S., Schiefer, N., Maxwell, T., Joseph, N., Hatfield-Dodds, Z.,
Tamkin, A., Nguyen, K., ... Olah, C. (2023). \emph{Towards
monosemanticity: Decomposing language models with dictionary learning}.
Transformer Circuits Thread.
\url{https://transformer-circuits.pub/2023/monosemantic-features/index.html}

Bright, J., Enock, F. E., Esnaashari, S., Francis, J., Hashem, Y., \&
Morgan, D. (2024). Generative AI is already widespread in the public
sector: Evidence from a survey of UK public sector professionals.
\emph{Digital Government: Research and Practice}.
\url{https://doi.org/10.1145/3700140}

Buolamwini, J., \& Gebru, T. (2018). Gender shades: Intersectional
accuracy disparities in commercial gender classification. In S. A.
Friedler \& C. Wilson (Eds.), \emph{Proceedings of the 1st Conference on
Fairness, Accountability and Transparency} (Proceedings of Machine
Learning Research, Vol. 81, pp. 77--91). PMLR.
\url{https://proceedings.mlr.press/v81/buolamwini18a.html}

Burrell, J. (2016). How the machine 'thinks': Understanding opacity in
machine learning algorithms. \emph{Big Data \& Society, 3}(1).
\url{https://doi.org/10.1177/2053951715622512}

Chouldechova, A. (2017). Fair prediction with disparate impact: A study
of bias in recidivism prediction instruments. \emph{Big Data, 5}(2),
153--163. \url{https://doi.org/10.1089/big.2016.0047}

Citron, D. K., \& Pasquale, F. (2014). The scored society: Due process
for automated predictions. \emph{Washington Law Review, 89}(1), 1--33.
\url{https://digitalcommons.law.uw.edu/wlr/vol89/iss1/2}

College of Policing. (2025). \emph{Building AI-enabled tools and
systems: Authorised professional practice}.
\url{https://www.college.police.uk/guidance/building-ai-enabled-tools-and-systems}

Department for Science, Innovation and Technology. (2025a). \emph{AI
opportunities action plan}. GOV.UK.
\url{https://www.gov.uk/government/publications/ai-opportunities-action-plan/ai-opportunities-action-plan}

Department for Science, Innovation and Technology. (2025b). \emph{State
of digital government review}. GOV.UK.
\url{https://www.gov.uk/government/publications/state-of-digital-government-review/state-of-digital-government-review}

Dressel, J., \& Farid, H. (2018). The accuracy, fairness, and limits of
predicting recidivism. \emph{Science Advances, 4}(1), eaao5580.
\url{https://doi.org/10.1126/sciadv.aao5580}

Elhage, N., Hume, T., Olsson, C., Schiefer, N., Henighan, T., Kravec,
S., Hatfield-Dodds, Z., Lasenby, R., Drain, D., Chen, C., Grosse, R.,
McCandlish, S., Kaplan, J., Amodei, D., Wattenberg, M., \& Olah, C.
(2022). \emph{Toy models of superposition}. Transformer Circuits Thread.
\url{https://transformer-circuits.pub/2022/toy_model/index.html}

European Union. (2024). \emph{Regulation (EU) 2024/1689 of the European
Parliament and of the Council of 13 June 2024 laying down harmonised
rules on artificial intelligence (Artificial Intelligence Act)}.
Official Journal of the European Union, L 2024/1689.
\url{http://data.europa.eu/eli/reg/2024/1689/oj}

Gallegos, I. O., Rossi, R. A., Barrow, J., Tanjim, M. M., Kim, S.,
Dernoncourt, F., Yu, T., Zhang, R., \& Ahmed, N. K. (2024). Bias and
fairness in large language models: A survey. \emph{Computational
Linguistics, 50}(3), 1097--1179.
\url{https://doi.org/10.1162/coli_a_00524}

Ganguli, D., Lovitt, L., Kernion, J., Askell, A., Bai, Y., Kadavath, S.,
Mann, B., Perez, E., Schiefer, N., Ndousse, K., Jones, A., Bowman, S.,
Chen, A., Conerly, T., DasSarma, N., Drain, D., Elhage, N., El-Showk,
S., Fort, S., \ldots{} Clark, J. (2022). \emph{Red teaming language
models to reduce harms: Methods, scaling behaviors, and lessons
learned}. arXiv. \url{https://arxiv.org/abs/2209.07858}

Glazer, E., Erdil, E., Besiroglu, T., Chicharro, D., Chen, E., Gunning,
A., Falkman Olsson, C., Denain, J.-S., Ho, A., de Oliveira Santos, E.,
Järviniemi, O., Barnett, M., Sandler, R., Vrzala, M., Sevilla, J., Ren,
Q., Pratt, E., Levine, L., Barkley, G., \ldots{} Wildon, M. (2024).
\emph{FrontierMath: A benchmark for evaluating advanced mathematical
reasoning in AI}. arXiv. \url{https://arxiv.org/abs/2411.04872}

Guha, N., Nyarko, J., Ho, D. E., Ré, C., Chilton, A., Narayana, A.,
Chohlas-Wood, A., Peters, A., Waldon, B., Rockmore, D. N., Zambrano, D.,
Talisman, D., Hoque, E., Surani, F., Fagan, F., Sarfaty, G., Dickinson,
G. M., Porat, H., Hegland, J., \ldots{} Li, Z. (2023). LegalBench: A
collaboratively built benchmark for measuring legal reasoning in large
language models. In \emph{Advances in Neural Information Processing
Systems 36 (NeurIPS 2023), Datasets and Benchmarks Track}.
\url{https://arxiv.org/abs/2308.11462}

Hackenburg, K., Tappin, B. M., Hewitt, L., Saunders, E., Black, S., Lin,
H., Fist, C., Margetts, H., Rand, D. G., \& Summerfield, C. (2025). The
levers of political persuasion with conversational artificial
intelligence. \emph{Science, 390}(6777), Article eaea3884.
\url{https://doi.org/10.1126/science.aea3884}

Harcourt, B. E. (2007). \emph{Against prediction: Profiling, policing,
and punishing in an actuarial age}. University of Chicago Press.

Hofmann, V., Kalluri, P. R., Jurafsky, D., \& King, S. (2024). AI
generates covertly racist decisions about people based on their dialect.
\emph{Nature, 633(8028)}, 147--154.
\url{https://doi.org/10.1038/s41586-024-07856-5}

Home Office. (2026). \emph{PoliceAI to speed up investigations and fight
crime}. GOV.UK.
\url{https://www.gov.uk/government/news/policeai-to-speed-up-investigations-and-fight-crime}

Infocomm Media Development Authority \& Personal Data Protection
Commission. (2022). \emph{AI Verify: An AI governance testing framework
and toolkit}. IMDA/PDPC.
\url{https://www.imda.gov.sg/resources/press-releases-factsheets-and-speeches/press-releases/2022/sg-launches-worlds-first-ai-testing-framework-and-toolkit-to-promote-transparency}

INTERPOL \& UNICRI. (2024). \emph{Toolkit for responsible AI innovation
in law enforcement} (Rev. ed.).
\url{https://www.interpol.int/en/How-we-work/Innovation/Artificial-Intelligence-Toolkit}

Jacobs, A. Z., \& Wallach, H. (2021). Measurement and fairness. \emph{In
Proceedings of the 2021 ACM Conference on Fairness, Accountability, and
Transparency} (pp. 375--385). Association for Computing Machinery.
\url{https://doi.org/10.1145/3442188.3445901}

Japkowicz, N., \& Shah, M. (2011). \emph{Evaluating learning algorithms:
A classification perspective}. Cambridge University Press.

Jimenez, C. E., Yang, J., Wettig, A., Yao, S., Pei, K., Press, O., \&
Narasimhan, K. (2024). SWE-bench: Can language models resolve real-world
GitHub issues? In \emph{The Twelfth International Conference on Learning
Representations (ICLR 2024)}. \url{https://arxiv.org/abs/2310.06770}

Jordan, M. I., \& Mitchell, T. M. (2015). Machine learning: Trends,
perspectives, and prospects. \emph{Science}, 349(6245), 255--260.
https://doi.org/10.1126/science.aaa8415

Liang, P., Bommasani, R., Lee, T., Tsipras, D., Soylu, D., Yasunaga, M.,
Zhang, Y., Narayanan, D., Wu, Y., Kumar, A., Newman, B., Yuan, B., Yan,
B., Zhang, C., Cosgrove, C., Manning, C. D., Ré, C., Acosta-Navas, D.,
Hudson, D. A., \ldots{} Koreeda, Y. (2023). Holistic evaluation of
language models. \emph{Transactions on Machine Learning Research}.
\url{https://arxiv.org/abs/2211.09110}

Lum, K., \& Isaac, W. (2016). To predict and serve? \emph{Significance,
13}(5), 14--19. \url{https://doi.org/10.1111/j.1740-9713.2016.00960.x}

Lundberg, S. M., \& Lee, S.-I. (2017). A unified approach to
interpreting model predictions.~In \emph{Advances in Neural Information
Processing Systems 30}.

Mehrabi, N., Morstatter, F., Saxena, N., Lerman, K., \& Galstyan, A.
(2021). A survey on bias and fairness in machine learning. \emph{ACM
Computing Surveys, 54}(6), 1--35. \url{https://doi.org/10.1145/3457607}

Mialon, G., Fourrier, C., Swift, C., Wolf, T., LeCun, Y., \& Scialom, T.
(2024). GAIA: A benchmark for general AI assistants. In \emph{The
Twelfth International Conference on Learning Representations (ICLR
2024)}. \url{https://arxiv.org/abs/2311.12983}

Mitchell, M., Wu, S., Zaldivar, A., Barnes, P., Vasserman, L.,
Hutchinson, B., Spitzer, E., Raji, I. D., \& Gebru, T. (2019). Model
cards for model reporting. In \emph{Proceedings of the Conference on
Fairness, Accountability, and Transparency} (pp. 220--229). Association
for Computing Machinery. \url{https://doi.org/10.1145/3287560.3287596}

National Institute of Standards and Technology. (2023). \emph{Artificial
intelligence risk management framework (AI RMF 1.0)} (NIST AI 100--1).
U.S. Department of Commerce. \url{https://doi.org/10.6028/NIST.AI.100-1}

National Police Chiefs' Council. (2023). \emph{Covenant for using
artificial intelligence (AI) in policing}. Office of the Police Chief
Scientific Adviser.
\url{https://science.police.uk/site/assets/files/4682/ai_principles_1_1_1.pdf}

Organization for Economic Co-operation and Development. (2019).
\emph{Recommendation of the Council on artificial intelligence}
(OECD/LEGAL/0449).
\url{https://legalinstruments.oecd.org/en/instruments/OECD-LEGAL-0449}

Organization for Economic Co-operation and Development. (2024).
\emph{National AI policies \& strategies}. OECD.AI Policy Observatory.
Retrieved July 13, 2026, from
\url{https://oecd.ai/en/dashboards/overview}

Oswald, M., Calder, M., Paterson-Young, C., \& Dunkwu, J. (2025).
\emph{PROBabLE Futures and NPCC responsible AI checklist for policing}.
Northumbria University. \url{https://bit.ly/PROBabLEFuturesNPCC}

Oswald, M., Grace, J., Urwin, S., \& Barnes, G. C. (2018). Algorithmic
risk assessment policing models: Lessons from the Durham HART model and
'experimental' proportionality. \emph{Information \& Communications
Technology Law, 27}(2), 223--250.
\url{https://doi.org/10.1080/13600834.2018.1458455}

Personal Data Protection Commission \& Infocomm Media Development
Authority. (2020). \emph{Model artificial intelligence governance
framework} (2nd ed.). PDPC.
\url{https://www.pdpc.gov.sg/-/media/files/pdpc/pdf-files/resource-for-organisation/ai/sgmodelaigovframework2.pdf}

Policing Productivity Review. (2023). \emph{Policing Productivity
Review:} Improving outcomes for the public \emph{{[}Independent
review{]}}. Home Office.
\url{https://www.gov.uk/government/publications/policing-productivity-review}

\emph{R (Bridges) v Chief Constable of South Wales Police} {[}2020{]}
EWCA Civ 1058.

Raji, I. D., Bender, E. M., Paullada, A., Denton, E., \& Hanna, A.
(2021). AI and the everything in the whole wide world benchmark. In
\emph{Proceedings of the 35th Conference on Neural Information
Processing Systems (NeurIPS 2021) Datasets and Benchmarks Track (Round
2)}. \url{https://arxiv.org/abs/2111.15366}

Raji, I. D., Smart, A., White, R. N., Mitchell, M., Gebru, T.,
Hutchinson, B., Smith-Loud, J., Theron, D., \& Barnes, P. (2020).
Closing the AI accountability gap: Defining an end-to-end framework for
internal algorithmic auditing. In \emph{Proceedings of the 2020
Conference on Fairness, Accountability, and Transparency} (pp. 33--44).
Association for Computing Machinery.
\url{https://doi.org/10.1145/3351095.3372873}

Rein, D. (2025, August 12). Research update: Algorithmic vs. holistic
evaluation. \emph{METR}.
\url{https://metr.org/blog/2025-08-12-research-update-towards-reconciling-slowdown-with-time-horizons/}

Rein, D., Hou, B. L., Stickland, A. C., Petty, J., Pang, R. Y., Dirani,
J., Michael, J., \& Bowman, S. R. (2023). GPQA: A graduate-level
Google-proof Q\&A benchmark. arXiv.
\url{https://arxiv.org/abs/2311.12022}

Ribeiro, M. T., Singh, S., \& Guestrin, C. (2016). ``Why should I trust
you?'': Explaining the predictions of any classifier. In
\emph{Proceedings of the 22nd ACM SIGKDD International Conference on
Knowledge Discovery and Data Mining} (pp. 1135--1144). Association for
Computing Machinery. \url{https://doi.org/10.1145/2939672.2939778}

Richardson, R., Schultz, J. M., \& Crawford, K. (2019). Dirty data, bad
predictions: How civil rights violations impact police data, predictive
policing systems, and justice. \emph{New York University Law Review
Online, 94}, 192--233.
\url{https://www.nyulawreview.org/wp-content/uploads/2019/04/NYULawReview-94-Richardson-Schultz-Crawford.pdf}

Rudin, C. (2019). Stop explaining black box machine learning models for
high stakes decisions and use interpretable models instead. \emph{Nature
Machine Intelligence, 1}(5), 206--215.
\url{https://doi.org/10.1038/s42256-019-0048-x}

Singhal, K., Tu, T., Gottweis, J., Sayres, R., Wulczyn, E., Amin, M.,
Hou, L., Clark, K., Pfohl, S. R., Cole-Lewis, H., Neal, D., Rashid, Q.
M., Schaekermann, M., Wang, A., Dash, D., Chen, J. H., Shah, N. H.,
Lachgar, S., Mansfield, P. A., \ldots{} Natarajan, V. (2025). Toward
expert-level medical question answering with large language models.
\emph{Nature Medicine, 31}(3), 943--950.
\url{https://doi.org/10.1038/s41591-024-03423-7}

Turpin, M., Michael, J., Perez, E., \& Bowman, S. R. (2023). Language
models don't always say what they think: Unfaithful explanations in
chain-of-thought prompting. In \emph{Advances in Neural Information
Processing Systems 36} (pp. 74952--74965).
\url{https://arxiv.org/abs/2305.04388}

Wallach, H., Desai, M., Cooper, A. F., Wang, A., Atalla, C., Barocas,
S., Blodgett, S. L., Chouldechova, A., Corvi, E., Dow, P. A.,
Garcia-Gathright, J., Olteanu, A., Pangakis, N. J., Reed, S., Sheng, E.,
Vann, D., Wortman Vaughan, J., Vogel, M., Washington, H., \& Jacobs, A.
Z. (2025). Position: Evaluating generative AI systems is a social
science measurement challenge. In \emph{Proceedings of the 42nd
International Conference on Machine Learning}.
\url{https://openreview.net/forum?id=1ZC4RNjqzU}

Wei, J., Tay, Y., Bommasani, R., Raffel, C., Zoph, B., Borgeaud, S.,
Yogatama, D., Bosma, M., Zhou, D., Metzler, D., Chi, E. H., Hashimoto,
T., Vinyals, O., Liang, P., Dean, J., \& Fedus, W. (2022). Emergent
abilities of large language models. \emph{Transactions on Machine
Learning Research}. \url{https://openreview.net/forum?id=yzkSU5zdwD}

Weidinger, L., Rauh, M., Marchal, N., Manzini, A., Hendricks, L. A.,
Mateos-Garcia, J., Bergman, S., Kay, J., Griffin, C., Bariach, B.,
Gabriel, I., Rieser, V., \& Isaac, W. (2023). \emph{Sociotechnical
safety evaluation of generative AI systems}. arXiv.
\url{https://arxiv.org/abs/2310.11986}
\end{hangparas}

\end{document}